\documentclass[conference]{IEEEtran}
\IEEEoverridecommandlockouts
\usepackage{cite}
\usepackage{amsmath,amssymb,amsfonts}

\usepackage{hyperref}
\hypersetup{
    colorlinks=true,
    linkcolor=blue,
    filecolor=magenta,      
    urlcolor=cyan,
    pdftitle={Overleaf Example},
    pdfpagemode=FullScreen,
    }
\usepackage{graphicx}

\usepackage{algorithmic}
\usepackage{graphicx}
\usepackage{textcomp}
\usepackage{xcolor}

\usepackage{caption}
\usepackage{subcaption}
\usepackage{float}
\usepackage[htt]{hyphenat} % for text wrapping in \texttt sections
\usepackage{array}
\usepackage{arydshln}

\usepackage{svg}
\usepackage[capitalize]{cleveref}
\usepackage{pifont}% http://ctan.org/pkg/pifont
\usepackage{lipsum}  
\usepackage{diagbox}
\usepackage{multirow}
\usepackage[export]{adjustbox}

\usepackage{listings}

\crefname{section}{Sec.}{Secs.}
\Crefname{section}{Section}{Sections}
\Crefname{table}{Table}{Tables}
\crefname{table}{Tab.}{Tabs.}
\crefname{lstlisting}{List.}{Lists.}

\def\BibTeX{{\rm B\kern-.05em{\sc i\kern-.025em b}\kern-.08em
    T\kern-.1667em\lower.7ex\hbox{E}\kern-.125emX}}
\begin{document}

\title{Toward Accessible and Safe Live Streaming Using Distributed Content Filtering with MoQ}

\author{
    \IEEEauthorblockN{Andrew C. Freeman}
    \IEEEauthorblockA{Baylor University
    \\andrew\_freeman@baylor.edu}
}

\maketitle

\begin{abstract}
%100-150 words
Live video streaming is increasingly popular on social media platforms. With the growth of live streaming comes an increased need for robust content moderation to remove dangerous, illegal, or otherwise objectionable content. Whereas video on demand distribution enables offline content analysis, live streaming imposes restrictions on latency for both analysis and distribution. In this paper, we present extensions to the in-progress Media Over QUIC Transport protocol that enable real-time content moderation in one-to-many video live streams. Importantly, our solution removes only the video segments that contain objectionable content, allowing playback resumption as soon as the stream conforms to content policies again. Content analysis tasks may be transparently distributed to arbitrary client devices. We implement and evaluate our system in the context of light strobe removal for photosensitive viewers, finding that streaming clients experience an increased latency of only one group-of-pictures  duration.
\end{abstract}


\section{Introduction}
\label{sec:intro}

In 2023, live streaming constituted 18\% of all downstream Internet traffic in the Americas \cite{sandvine_2024_2024}. Although some of this traffic is in traditional media such as sports and news, live streaming content is dominated by social media platforms such as YouTube, Twitch, Facebook, Instagram, X, and TikTok. For example, TikTok reported that more than 100 million of its users created a live stream in 2024 \cite{tiktok_pte_ltd_celebrating_2025}. Whereas government bodies such as the Federal Communications Commission regulate what content is permissible on broadcast television, there is much less oversight for Internet-based platforms. Indeed, the massive scale of user-generated live stream content is impossible to police through manual means.

In particular, video streaming platforms seek to detect and mitigate risks such as explicit content, violence, illegal acts, and photosensitivity triggers. Many of these risks can be detected through existing offline analysis pipelines for video on demand (VoD) content. Then, a platform may simply block the publication or visibility of an entire video if it contains objectionable content. In the live context, however, a video platform must analyze a stream in real time, blocking it only when the objectionable content \textit{begins}.

Dominant streaming solutions such as HTTP Live Streaming (HLS) and Dynamic Adaptive Streaming over HTTP (DASH) have ``Low-Latency'' extensions for live streaming in LL-HLS and LL-DASH, respectively. These existing systems are generally stateless, meaning that clients determine the data they receive by issuing specific HTTP GET requests. The servers likewise cannot easily choose which data to make available for any individual client. 

Meanwhile, Media Over QUIC (MoQ) is a protocol draft aimed at improving the latency of live video delivery. MoQ is stateful, giving servers the power to determine exactly what data any particular client may receive. We argue that stateful streaming with MoQ can greatly improve the flexibility of content moderation tools, allowing content to be selectively blocked with user-specific criteria. For example, video segments depicting alcohol or tobacco use may be blocked for underage viewers, but still be transmitted to adult viewers. 

In this paper, we first examine the current design of MoQ for live video streaming. We then propose extensions for dynamic content filtering with distributed content analysis. We evaluate our system implementation with a toy ``light strobe'' detection application for photosensitive viewers. We demonstrate that the analysis step may delay the delivery of approved content by only the duration of a single group-of-pictures. Finally, we discuss the remaining steps to to incorporate rate adaptation and server-driven task distribution.%Our source code will be released upon publication.%\footnote{We have included a video demonstration as supplementary material for the reviewers.}.
% Furthermore, the analysis may equivalently run either on a delivery server or on an end client, with no changes necessary for the transport protocol. 

% We envision a stateful live streaming system 

% TODO: Introduce and explain the acronyms: ULVS, VoD, MoQ, GOP.

% Idea works by specifying an additional \textit{control message} AND modifying the subscribeUpdate message to include whether or not a subscriber is an analyzer.

\section{Related Work}

\subsection{HTTP Adaptive Streaming}

As noted above, the most common video streaming protocols today are HLS and DASH, falling under the umbrella of HTTP Adaptive Streaming (HAS) \cite{bentaleb_toward_2025}. These protocols operate by dividing a source video file into temporal \textit{segments}, and these segments are catalogued in a \textit{manifest} (or ``playlist'') file \cite{altamimi_client-server_2019}. A segment typically maps to a group of pictures (GOP), an independently decodable time range of the video. The manifest and source video files are made available on an HTTP server. To receive video data, a streaming client first issues a GET request for the manifest. The manifest provides the necessary information for subsequent GET request to retrieve each desired segment of the video. These protocols support \textit{adaptation} through indexing multiple representations of a video at different bitrates. Based on network congestion, a client may lower or increase the bitrate by simply changing the target representation in the GET request for the next segment.

Live streaming variants of these protocols (LL-HLS and LL-DASH) operate on much the same principle, but with some modifications for low-latency updates. Video segments are further divided into multiple ``chunks,'' which are smaller file representations corresponding to a handful of image frames, rather than an entire GOP \cite{bentaleb_low_2022,bentaleb_toward_2025}. This chunking process allows the client to retrieve (and begin decoding) a new GOP before the entire GOP has been written to the HTTP server. Live-oriented HAS protocols increase the communication overhead compared to VoD, because clients must issue more frequent GET requests to retrieve dynamic manifest updates and sub-segment chunks. 

\subsection{Media Over QUIC (MoQ)}
MoQ is a work-in-progress protocol being developed through the Internet Engineering Task Force (IETF). It primarily aims to improve the latency and bandwidth requirements for live stream video delivery \cite{gurel_media-over-quic_2024}. MoQ Transport (MoQT) is the core document in the MoQ protocol draft \cite{curley_media_2025-1}. The MoQT specification defines the data entities, header formats, and control signals of the transport protocol. Here, we offer a brief summary of the components from MoQT that are relevant to this work. Although MoQT is a generic transport protocol, for this paper we use terminology specific to video streaming.

MoQ uses a publisher/subscriber model. Data are communicated from a \textit{publisher} to a \textit{relay} \cite{bentaleb_toward_2025,curley_media_2025-1}. A relay then manages \textit{sessions} for publishers and \textit{subscriber(s)}, forwarding the data to each subscriber as they are made available. Connections are maintained through QUIC or WebTransport streams, unlocking push-based media delivery. The protocol draft does not specify how a relay may or may not allow interaction between different subscriber sessions.

Stream adaptation is made possible through the use of \textit{Tracks}, which contain independent data streams. A client may, for example, subscribe to a high-bitrate video Track when it has a strong connection, then replace its subscription with a low-bitrate Track when it encounters network congestion. Tracks carry \textit{Groups}, which are typically independently decodable units of data. In video streaming, a Group commonly maps to a GOP in an encoded video. A Group consists of one or more \textit{Objects}, which map to individual video frames. Subscribers can set the \textit{priority} of each subscription, which determines how the relay will deliver data during periods of congestion. For example, a client may set an audio Track subscription to have a higher priority than a video Track. The WARP stream format aims to standardize the mechanisms for MoQ-based video adaptation \cite{law_warp_2025}.

% - How it works, what the major components are
% - Publisher
% - Relay
% - Subscriber
% - Sessions
% - Control messages
% - Objects
% - Lack of inter-session control
% - Latency stuff?

% Talk about how the specific sensitivity of each client could be tuned

\subsection{Content Analysis and Filtering}
% - What content should be detected and blocked?
% - Related work on detecting strobe problems in particular

Objectionable video content can fall into three classes: that which poses a risk to health; that which is illegal; and that which is offensive. Here, we discuss some prior work on detecting instances of each of these three classes in video.

\subsubsection{Health Risks}

The primary health-related issue with video content is the potential to cause seizures. There are more than 100,000 Americans with ``photosensitive epilepsy'' (PSE), a condition that increases the risk for seizures when individuals are subjected to certain light stimuli \cite{erba_shedding_2006}. These stimuli include rapidly flashing lights or colors and high-contrast patterns \cite{erba_shedding_2006,harding_photosensitive_2010,jordan_international_2024}. There are several guidelines for mitigating PSE risks in video content production, and these guidelines have driven the development of both direct and learned methods for PSE risk detection  \cite{jordan_international_2024,vu_access_2011,barbu_deep_2020}. Previous works mainly perform PSE risk detection offline, and risks are mitigated by modifying a video itself (e.g., by reducing contrast) \cite{jordan_international_2024}. In the context of HAS, such methods would require transcoding to produce a PSE-safe video representation. To our knowledge, no social media platform has implemented PSE-safe transcoding in their video streaming pipelines. In 2020, however, TikTok introduced an accessibility option that disables the playback of videos determined to be PSE risks through \cite{goodman_making_2020}. This feature appears to apply only to VoD, rather than TikTok's live service.

\subsubsection{Illegal Content}

Some filtering topics are considered illegal in nearly all jurisdictions, including depictions of child exploitation and sexual assault, and the distribution of pirated materials \cite{sartor_impact_2020}. Social media platforms leverage a combination of automated analysis, human moderators, and user reporting to police their video libraries for illegal content \cite{sartor_impact_2020}. 
% For example YouTube's Content ID tool compares audio and video content to a database of copyrighted materials, removing a video if it is found to contain such materials without
However, one should consider that the illegality of content may depend on the age and location of the viewer. For example, a country may want to prohibit the depiction of smoking or alcohol usage for minors. To support such blocking in traditional adaptive streaming systems, a platform must block access to the stream data at the application layer. For large-scale live streaming, the computational burden makes real-time content analysis difficult. Platforms such as Twitch thus rely heavily on human moderation for these streams \cite{cai_coordination_2022}. If a live stream is deemed to contain any illegal content, it is wholly blocked for all users.

\subsubsection{Offensive Content}

Finally, we shift our attention to content that is not necessarily illegal, but may be offensive to some users. Social media platforms often impose their own restrictions, such as prohbiting violence, drugs, gambling, or pornography \cite{valle_content-based_2011,shah_content_2021}. Again, in HAS live streaming, moderation of these topics is not done in real time. If a stream publisher begins filming offensive content during a stream, several minutes may pass before a stream is blocked by the platform's moderators (if it is blocked at all). 

Furthermore, there is no mechanism on social video platforms for individual users to dictate which specific topics they want to block in live streams. This is particularly an issue for parental control, as different sets of parents may have dramatically different views on what content is appropriate for their children.

% \subsection{Traditional Content Filtering}

\section{Motivation}\label{sec:motivation}
% [][What makes live video in particular so much more difficult?][Maybe move this before related work?]

We envision a live streaming system where each end user may choose precisely what content they want to block from their received stream. This system should support filtering a variety of user-selected content \textit{categories}, support distributed processing, and have a minimal impact on stream latency. If fully realized, it can bring about improvements to stream accessibility, safety, and personalization for users worldwide.

With traditional HAS protocols, an analysis pipeline can only be deployed on the publisher or the live stream ingest server. The server could continually add metadata to the manifest file (e.g., with DASH EventStream tags) to indicate where certain content categories are (or are not) detected.  The major downside to this approach is the increase in latency for all clients. For each new segment, every client fetches the updated manifest file. The client then would check to see if all of its filter categories have completed analysis and been approved. If some filter categories are still under analysis, the client must poll the server continually until they are complete. These polling requests, however, will invariably increase the end-to-end latency: either the server will slow down due to the frequent requests, or request rates will be throttled. Alternatively, the ISO-BMFF video fragments themselves could have this metadata appended in an appropriate format (e.g., Event Message boxes for DASH). To avoid repeated client requests, the segments would only be made available in the manifest once all analysis tasks have completed. Again, however, some analyses may take substantial time, and may not be relevant for all consumers. The overhead of pull-based media delivery thus makes such a system impractical. Furthermore, we cannot adequately harness the computational resources of client devices in a typical HAS system. If a client devices analyzes a video segment, it cannot disseminate the result to other streaming clients.

We argue instead that \textit{push}-based delivery can overcome these issues by minimizing overhead, reducing latency, and increasing the granularity of user personalization.

% [][TODO: Insert a figure illustrating this process]

% We argue that the present difficulty of live stream content filtering is largely a symptom of the stateless, pull-based media delivery of HAS protocols. 

% [What is our goal? Accessibility, safety, and personalized control]

\section{System Design}

\begin{figure*}
  \centering
    \includegraphics[width=\linewidth]{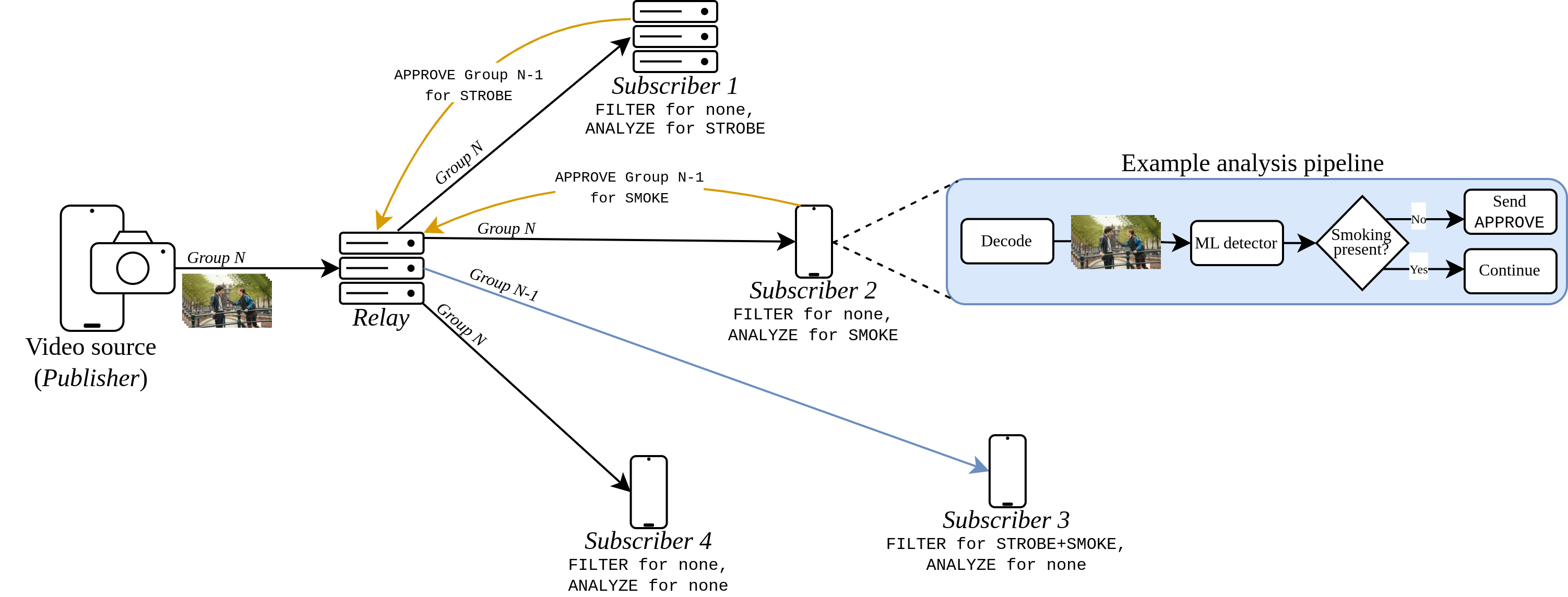}
    \caption{Overall diagram of our distributed content filtering system. Analysis tasks can transparently be distributed to both servers (e.g., Subscriber 1) and user devices (e.g., Subscriber 2) with the same MoQ mechanisms.}
    \label{fig:system_diagram}
\end{figure*}

We propose modifications to a MoQ live streaming system to enable dynamic content filtering. One or more susbcribers are designated as ``analyzers.'' An analyzer client decodes each Group it receives, and runs the frames through a computer vision analysis pipeline. The goal of the analysis is to determine if the Group contains any material that should be blocked for other subscribers. If the analyzer determines that the Group does \textit{not} contain any objectionable material, it notifies the relay that the Group is acceptable. When the relay receives this message, it then transmits that Group to all downstream subscribers that are awaiting the approval. We illustrate the system in \cref{fig:system_diagram}.

\subsection{Subscription Messages}

% To coordinate the distribution of analysis tasks and Group delivery, we propose the following new control messages for MoQT.

A client initiates a MoQ subscription by issuing a \texttt{SUBSCRIBE} control message to the relay, containing information such as the track identifiers and the priority \cite{curley_media_2025-1}. A client may modify an existing subscription by issuing a \texttt{SUBSCRIBE\_UPDATE} control message. In both cases, the draft specification allows for a number of optional ``parameters'' at the end of these messages for additional functionality. For example, a client can include a \texttt{DELIVERY\_TIMEOUT} parameter to set the maximum latency between an Object arriving at the relay and being sent to the subscriber.

For our system, we introduce two new MoQT parameters: \texttt{ANALYZE} and \texttt{FILTER}. Both parameters include a variable number of ``categories.'' A subscriber issues an \texttt{ANALYZE} parameter in a \texttt{SUBSCRIBE}-based message if it wants to perform certain content analyses on the received stream. If the relay accepts this request, then that subscriber is referred to as an ``analyzer'' for those categories of analysis. Similarly, a subscriber issues a \texttt{FILTER} parameter if it wants to only receive a Group once it has been approved by an analyzer for the requested categories. The only difference between \texttt{ANALYZE} and \texttt{FILTER} is the parameter Type. Both may contain a variable number of Categories, such as \texttt{STROBE}, \texttt{SMOKING}, and \texttt{ALCOHOL}.  We detail the structure of these parameters in \cref{lst:analyze} below, where ``(i)'' indicates that the data type is a variable-length integer, as in the protocol draft \cite{curley_media_2025-1}.

\begin{lstlisting}[caption={Format for the \texttt{ANALYZE} and \texttt{FILTER} parameters. If a client \texttt{ANALYZES} some Categories, it will examine each Group for the presence of those materials. If a client \texttt{FILTERS} some Categories, it will only receive a Group if it does not contain any of those materials.},captionpos=b,label=lst:analyze]
ANALYZE/FILTER Parameter {
    Parameter Type (i) = 0x05/0x06
    Parameter Length (i)
    Parameter Value = Categories {
        Categories Length (i)
        Number of Categories (i)
        [
            Category Type (i),
            Category Type (i),
            ...
        ]
    }
}
\end{lstlisting}

% For this early work, we only considered a single analysis application.

% TODO: Talk about how to orchestrate MULTIPLE content categories being optionally blocked. Distribute analysis for one category to each analyzer client.

% TODO: Talk about our assumptions with group sequence numbers increasing by 1 each time, and only in order

% TODO: Talk about what the relay must do to prevent deadlock if there are conflicting analyze and filter messages (both subscribers waiting on each other) --> Priority is given to the first one, some update messages might be ignored. 

\subsection{Group Approval Messages}

Once an analyzer subscriber has finished processing a Group, it issues an \texttt{APPROVE} control message. This message carries the subscription and Group identifiers (IDs), and the Categories that have been approved. If the subscriber is an analyzer for multiple Categories, it is possible that only a subset of them may be \texttt{APPROVED}. If there are \textit{no} approved Categories, the subscriber may avoid sending an \texttt{APPROVE} message entirely. We detail the format of this control message in \cref{lst:approve}.

\begin{lstlisting}[caption={Format for the \texttt{APPROVE} message. An analyzer client can report on one or more Categories, and an \texttt{APPROVE} message indicates that the Group does not contain the material conveyed by those Categories.},captionpos=b,label=lst:approve]
APPROVE Message {
    Type (i) = 0x41,
     Length (i),
     Subscribe ID (i),
     Group ID (i),
     Categories {
        Categories Length (i)
        Number of Categories (i)
        [
            Category Type (i),
            Category Type (i),
            ...
        ]
    }
}
\end{lstlisting}

\subsection{Session Management}

The relay is responsible for the coordination of various analyzer clients and the appropriate delivery of Groups. The relay sends a Group to a subscriber if and only if all of the subscriber's \texttt{FILTER} Categories have been \texttt{APPROVED} for that group by other subscribers. A subscriber should \textit{not} have both \texttt{ANALYZE} and \texttt{FILTER} mechanisms enabled; otherwise, it would not be able to analyze groups that are blocked by its filter Category. \cref{fig:system_diagram} demonstrates the additional latency incurred through \texttt{FILTERING}: Subscriber 3 must wait for Group $N-1$ to be analyzed by both Subscriber 1 and Subscriber 2. Since these processes may run concurrently, Subscriber 3 incurs the latency of whichever \texttt{ANALYZE} connection is the slowest. Meanwhile, the newest group in the source live stream is Group $N$, which is one GOP duration ahead.

\section{Implementation}

\subsection{Implementation Details}

As the basis for our system, we used the \texttt{moq-rs}\footnote{https://github.com/kixelated/moq-rs} Rust repository (commit ID 1d895c7). Although this repository technically tracks a fork of the core MoQ draft \cite{curley_media_2025}, the fundamental mechanisms we employed are not significantly different from the mainline protocol, and our proposed system may be implemented for other reference software in the future.

\begin{figure*}
  % \centering
    \includegraphics[width=0.9\linewidth]{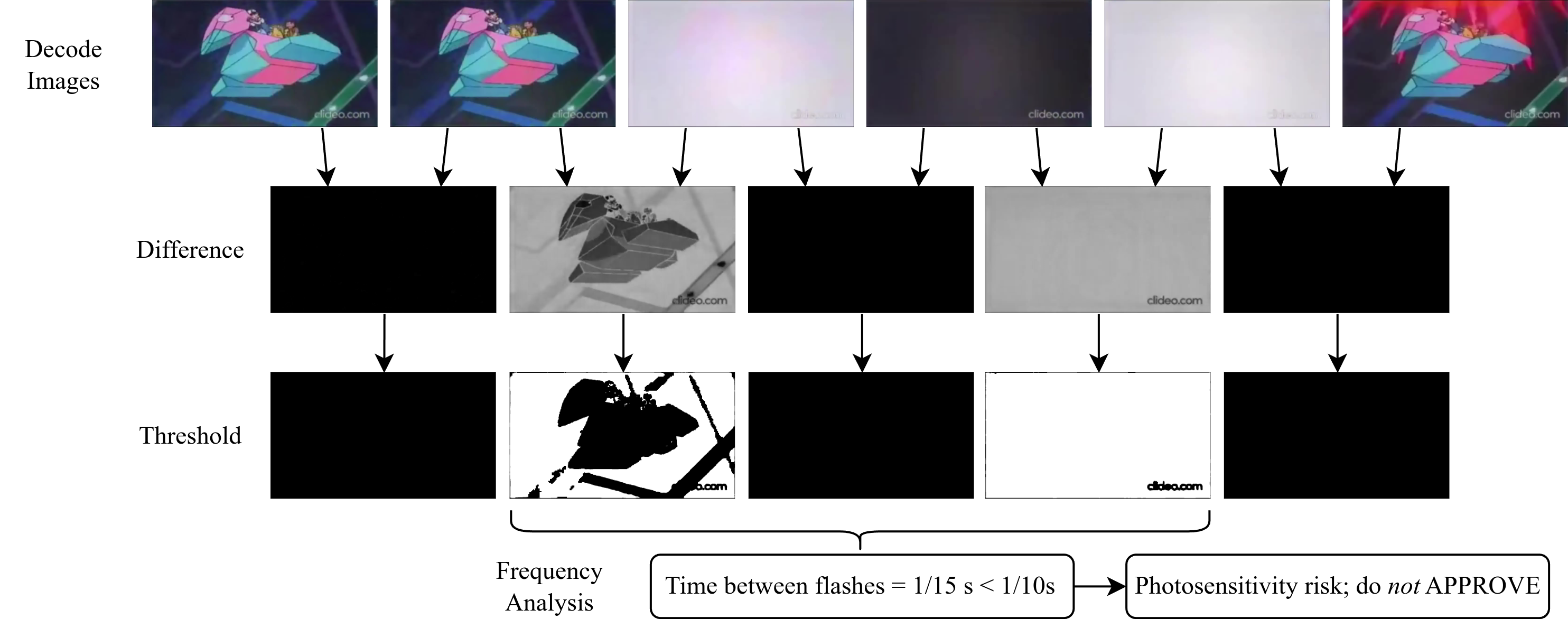}
    \caption{Example of our strobe detection application on the infamous Pokémon Episode 38, which was reported to have caused hundreds of Japanese children to experience epileptic seizures \cite{ishida_photosensitive_1998}. In our system, this flashing strobe sequence can be filtered so that photosensitive viewers are not exposed to risk. }
    \label{fig:pokemon}
    \hspace{0.1\linewidth}
\end{figure*}

As an example analysis application for this work, we implemented a toy light strobe detection system in the Web player using Rust for WebAssembly. Our application takes a uniform pixel sample of the luminosity (Y) channel of two consecutive YUV image frames. It evaluates the brightness difference of these samples, and compares it to a tunable threshold value. If a substantial portion of the samples have increased in brightness beyond the threshold, the frame is considered to be a significant brightness change. If there are at least two such change images during a short period of time (representing a strobe interval of at least 10 Hz), then we classify the Group as a strobe risk. Otherwise, the subscriber issues an \texttt{APPROVE} message for the Group, indicating that it does not pose a substantial risk for photosensitive viewers. \cref{fig:pokemon} illustrates the flow of this analysis pipeline. In our web player, we included an HTML button toggle a subscriber's status as an analyzer. When a user presses the button, the client sends a \texttt{SUBSCRIBE\_UPDATE} control message with the appropriate \texttt{ANALYZE} parameter.

We amended the transport reference code to appropriately manage the Group delivery. Rather than deliver newly-received Groups to all subscribers at once, our relay delivers them first to analyzer subscribers that have no \texttt{FILTER} Categories enabled. When an \texttt{APPROVE} message is received on an analyzer session, we forward it as a broadcast message on a Rust-based ``multiple producer, multiple consumer'' communication channel. All other active sessions on the relay wait to receive these messages, tracking which of their \texttt{FILTER} Categories have been \texttt{APPROVED}. Once all the Categories have been \texttt{APPROVED} for the given Group, the session finally delivers the Group to the subscriber. For simplicity, we assume that the Group IDs received by the relay are monotonically increasing by one. Therefore, if a session receives the relevant \texttt{APPROVE} messages for some Group $N$, but does not receive all the \texttt{APPROVE} messages for Group $N-1$, it knows that Group $N-1$ has been rejected for some Category. This Group is silently skipped, without informing the subscriber. The \texttt{APPROVED} Groups are buffered in a queue for each session, to accommodate variations in latency.

\subsection{Latency Analysis}

An important factor to consider in the deployment of a live filtering system is the overall latency for end users. We define the maximum latency for subscriber $y$ as 

\begin{multline}\label{eq:latency}
    L(y) = p + \max{\{ R(x) : x \in C\}} + \max{\{ F(x) : x \in C\}} \\+ R(y) + \max(G), 
\end{multline}

where $p$ gives the Group latency from the publisher to the relay, $R$ gives the Group latency from the relay to a subscriber, $C$ is the set of Categories that subscriber $y$ is filtering, $F$ gives the \texttt{APPROVE} message latency from a subscriber to the relay, and $\max(G)$ is the longest duration of any GOP in the published video stream. If subscriber $y$ maintains a constant playback speed, then a latency increase in any one of these components may cause video buffering events in the player. In practice, $G$ will typically not increase beyond a small maximum of one to two seconds, based on common recommendations for live stream encoding parameters \cite{hajihoseini_optimizing_2021}. Furthermore, it is crucial that any client-side analysis application can run in real time alongside the video playback. That is, the analysis duration for each GOP must be less than its playback duration. Otherwise, latency would continually increase for all downstream clients. We note that diverse applications may process different numbers of frames in each GOP, as needed.

We emphasize that our system unlocks a novel \textbf{user-tunable tradeoff between latency and content safety}. As noted in \cref{sec:motivation}, push-based streaming systems with content analysis pipelines will uniformly increase the latency for \textit{all} users, including those who do not wish to filter any content. In our system, in contrast, such users can receive the live stream with minimal latency, while \textit{only} the users with filter categories enabled will see an increase in end-to-end latency.

\section{Results}

We tested our implementation on a live webcam feed published to two separate web clients on a local machine. The GOP duration of the webcam stream was set to 1 second, and each GOP was organized into a separate Group. One client was set to \texttt{ANALYZE} light strobes, while the other was set to \texttt{FILTER} light strobes. Based on timestamp logs, we found that the filtered subscriber consistently experienced 994-1005 milliseconds of additional latency compared to the analyzer, approximately matching the GOP duration. We then introduced a strobing light impulse to the camera feed, observing that the playback on the filtered subscriber appeared to pause. In the background, the client stopped receiving new Groups until the impulse was removed.

The setup used for these experiments mirrors what we expect to be a common arrangement for these systems. That is, all content analysis is done on a server to minimize end-to-end latency. We then see less than 1 millisecond of propagation delay for both a Group to reach the subscriber, and for an \texttt{APPROVE} message to reach the relay and propagate to other sessions.

We emphasize, however, that the analysis and control mechanisms run on top of a standard MoQ subscriber. Therefore, we can trivially \textit{distribute} the analysis load to concurrent processes or even on the devices of end users. This can result in more energy-efficient and cost-effective streaming solutions when users have relaxed latency requirements. 
% [][TODO: Point to the figure throughout this description.]

% Secondly, we measured the impact of network congestion on $F$, the latency for \texttt{APPROVE} message delivery (\cref{eq:latency}). We used Google Chrome's network throttling tools with the preconfigured profiles Fast 4G, Slow 4G, and 3G to simulate bandwidth and latency limitations. For each profile, we published video streams at various bitrates and measured the average \texttt{APPROVE} latency, $F$.

% At the start of the live stream, both clients maintained synchrony with the source camera feed. 

\section{Future Work}

There is substantial room to improve our live stream filtering system. Chiefly, we have not yet established how our analysis and Group approval mechanisms will operate in an adpative streaming context. For example, an \texttt{APPROVE} message for Group $N$ should be distributed to all sessions subscribed to \textit{any} video track from the original publisher. That is, the presence of some objectionable content at one bitrate and resolution should preclude its dissemination at all other bitrates and resolutions. Significant improvements to the MoQ reference software will be necessary for this effort. The ongoing efforts to develop the MoQ WARP stream format will help elucidate the remaining necessary work \cite{law_warp_2025}.

Secondly, there is currently no mechanism to push analysis work onto arbitrary subscribers. For this, we suggest that a relay can send a \texttt{SUBSCRIBE\_UPDATE} message \textit{to} a subscriber with the \texttt{ANALYZE} parameter present. However, a relay-sourced \texttt{SUBSCRIBE\_UPDATE} message is currently an undefined behavior in the protocol draft \cite{curley_media_2025-1}. Additionally, the client and relay will need to exchange information about their supported analysis Categories during the session setup. This information may be added as parameters in the \texttt{CLIENT\_SETUP} and \texttt{SERVER\_SETUP} messages. In a real-world deployment, the relay should verify that client-side analysis code has not been modified, to prevent malicious attacks on the integrity of Group approvals. This aspect of the system has not yet been specified and will be a crucial area of further research.

Finally, we will solicit community feedback on our proposed system. Where appropriate, we will submit change requests to the current MoQT draft. For the components of our system that fall outside of MoQT, we will create a draft proposal with the IETF. This proposal will seek to standardize the methods for distributing and validating analysis applications to client players, the wire format and behavior of our custom parameters and messages, and the dynamic distribution of analysis tasks among several clients.

\section{Conclusion}

As more and more people engage with live video stream content, we aim to develop novel methods to improve the experience of everyday users. We proposed a novel system wherein content analysis can be transparently distributed among several processes and devices. Leveraging the lightweight MoQ Transport protocol, the distribution process can add as little as one GOP duration of latency if analyses are executed in real time aboard the MoQ relay server. Motivated by the dirth of accessibility options for users with PSE, we developed a toy ``light strobe'' application to demonstrate the efficacy of our system implementation. Finally, we discussed how users may specify their individual preferences for content filtering, and noted that latency will scale with the number of filters applied. As the industry contends with demands for increased accessibility, safety, personalization, and speed in live stream moderation, our work merely represents a first step. Future work will focus on the necessary changes for standardization, stream adaptation, and deployment.

\bibliographystyle{IEEEbib}
\bibliography{references}

\end{document}